\documentclass{PoS}
\usepackage{caption}
\usepackage{subcaption}

\title{Open Heavy Flavour: Experimental summary}

\ShortTitle{Open HF: experimental summary}

\author{\speaker{Deepa Thomas}
\thanks{This work was supported by U.S. Department of Energy Office of Science under contract number $\rm{DE-SC0013391}$.}\\
        The University of Texas at Austin\\
        E-mail: \email{deepa.thomas@cern.ch}}

\abstract{In this paper I will review a few of the latest experimental measurements of heavy-flavour hadrons presented at the Hard Probes 2018 conference. Results from experiments both at RHIC and at the LHC will be discussed. I will present some of the open questions that still need to be addressed with heavy quarks in small collision systems and in A-A collisions, to better understand the properties of the QGP produced in heavy-ion collisions. I will discuss some of the open heavy-flavour measurements designed to give insights into these questions. }

\FullConference{International Conference on Hard and Electromagnetic Probes of High-Energy Nuclear Collisions\\
		30 September - 5 October 2018\\
		Aix-Les-Bains, Savoie, France}

\begin{document}
\section{Introduction}
Heavy quarks (charm and beauty) are excellent probes to study the QCD medium formed in high-energy hadronic collisions. Due to their large masses, they are produced in hard scattering processes in the initial stages of the collisions with large $Q^{2}$ values~\cite{Andronic:2015wma}; thus their production is calculable within the framework of perturbative-QCD down to low transverse momentum ($p_{\rm{T}}$). In nucleus-nucleus collisions, heavy quarks are a sensitive probe of the hot, strongly-interacting medium, as their production time scales are shorter than the QGP thermalisation time~\cite{Liu:2012ax}, and their numbers are conserved in the full evolution of the medium. As heavy quarks propagate through the QGP, they undergo elastic (collisional) and inelastic (radiational) collisions and lose some of their momentum, and are thus sensitive to the transport properties of the medium. Understanding the medium-induced effects requires an accurate study of the cold nuclear-matter (CNM) effects in the initial and final stages of the collision, which modify the production of heavy quarks in nuclear collisions relative to pp collisions. The CNM effects are studied with heavy-flavour measurements in p-A and d-A collisions. Heavy quarks are studied with heavy-flavour hadrons, which are measured either via full reconstruction of their decays or with semi-inclusive decay daughters. 

Experiments at RHIC and at the LHC have provided a wealth of heavy-flavour hadron measurements in pp collisions which are differential in $p_{\rm{T}}$ and in rapidity ($y$), and which are well described by pQCD calculations within uncertainties~\cite{Andronic:2015wma}. The nuclear modification factor in  minimum-bias p-A collisions ($R_{\rm{pA}}$) for leptons from decay of heavy-flavour hadrons and D mesons, measured at mid- and at large-rapidities at RHIC and at the LHC, shows no large suppression of the yield at high  $p_{\rm{T}}$. These measurements are well described within uncertainties by models that include CNM effects~\cite{Acharya:2017hdv,Adam:2016ich}.  We also have quite precise measurements of nuclear modification factor ($R_{\rm{AA}}$), azimuthal anisotropy ($v_2$) for heavy-flavour decay leptons and D mesons at RHIC and LHC energies. The $R_{\rm{AA}}$ measurements show a large suppression of yields at high-$p_{\rm{T}}$ indicating that charm quarks interact strongly with the QGP and lose energy~\cite{Acharya:2018hre}. The $R_{\rm{AA}}$ of D mesons is also found to be similar to that of pions at high $p_{\rm{T}}$. Measurements of jets containing B hadrons at LHC also show a similar suppression pattern as that of inclusive jets at high $p_{\rm{T}}$~\cite{Sirunyan:2018jju}. A positive $v_2$ for D mesons has also been measured, suggesting that low-$p_{\rm{T}}$ charm quarks participate in the collective motion of the medium~\cite{Adamczyk:2017xur}. A qualitative description of  $R_{\rm{AA}}$ and $v_2$ can be obtained using models that include hydrodynamic expansion of the QGP with heavy quarks undergoing collisional and radiative energy loss, with hadronization via quark recombination in addition to vacuum fragmentation~\cite{Acharya:2018hre}. While these models provide a fair description of the data in some $p_{\rm{T}}$ regions, it is still a challenge to provide a simultaneous description of  $R_{\rm{AA}}$ and $v_2$ in the full $p_{\rm{T}}$ range.  

There are still many open questions on the production of heavy-flavour hadrons as a function of event multiplicity in small systems (pp and p-A collisions). In A-A collisions, a better understanding of the mass and flavour dependent energy loss of heavy quarks in the QGP is needed. Further studies on the hadronization and baryon production mechanisms in pp, p-A and A-A collisions are necessary. Experiments are also investigating the modification of the jet-structure and its kinematics in A-A collisions. A better description of the initial conditions in heavy-ion collisions is needed. 

In this paper, I will present some of the open heavy-flavour measurements presented at Hard Probes 2018 that could give some insights towards answering the open questions mentioned above, with the aim to further understand the properties of the QCD medium. 

\section{Heavy-flavour production in p-Pb and p/d-A collisions} 
\subsection*{Centrality dependent heavy-flavour production}
\begin{figure}
	\centering
\includegraphics[scale=0.28]{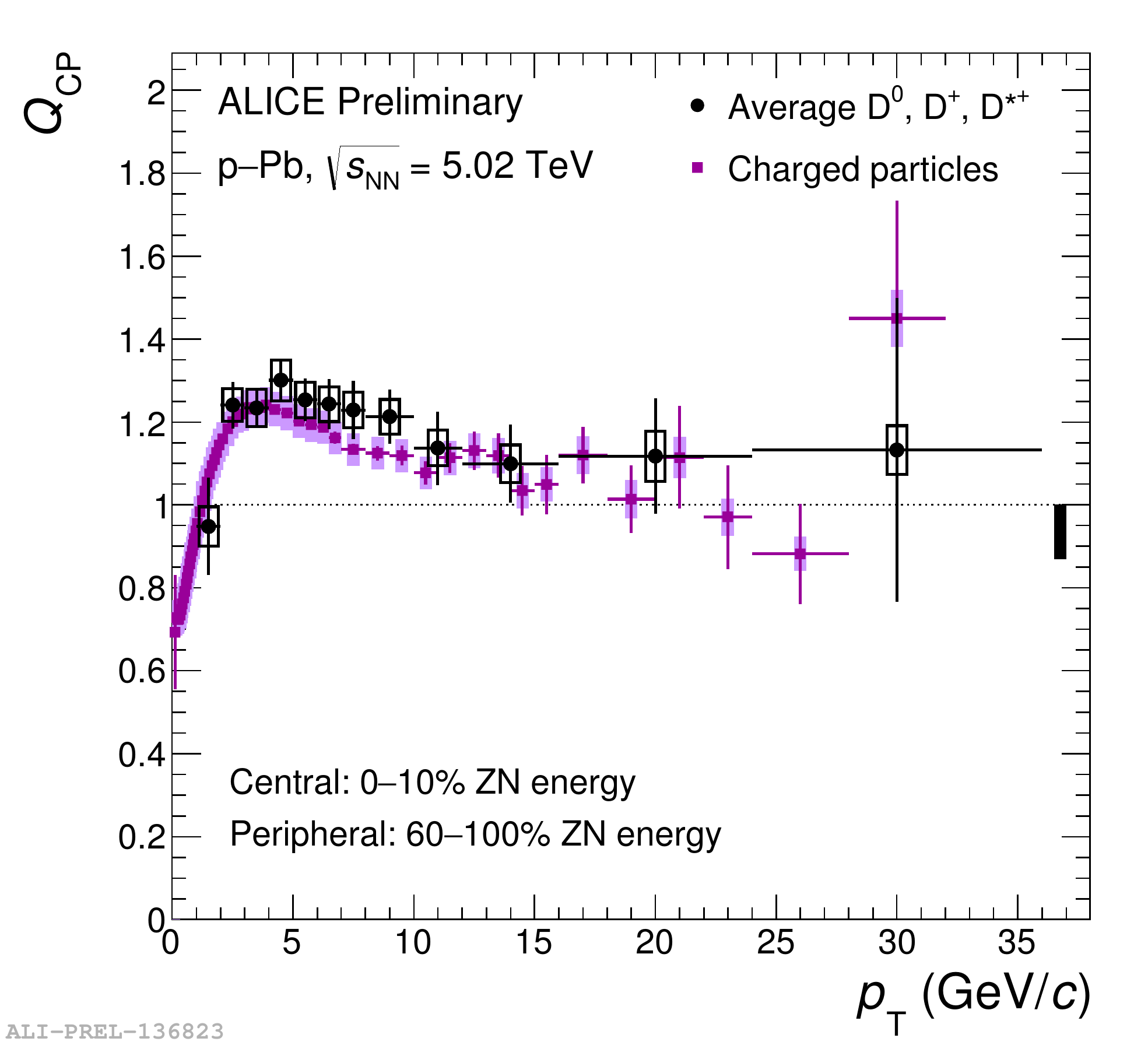}
\caption{The ratio of D meson cross-section in $0-10\%$ central to $60-100\%$ central p-Pb collisions at $\sqrt{s_{\rm{NN}}}=5.02$ TeV.}
\label{fig:DQcp}
\end{figure}

To study the multiplicity dependent production of heavy quarks, ALICE measured the D meson (average of $\rm{D}^{0}$, $\rm{D}^{+}$ and $\rm{D}^{*+}$) cross-section as a function of centrality at mid-rapidity in p-Pb collisions at $\sqrt{s_{\rm{NN}}}=5.02$ TeV. The ratio of the D meson cross-section in $0-10\%$ central to $60-~100\%$ central p-Pb collisions, referred to as $Q_{\rm{CP}}$, was obtained as shown in Figure~\ref{fig:DQcp}. The measured $Q_{\rm{CP}}$ shows an enhancement of D meson yield at low $p_{\rm{T}}$ in central p-Pb collisions compared to peripheral collisions, similar to what is observed for charged particles.  The observed enhancement for D mesons at low $p_{\rm{T}}$ is qualitatively similar to what was observed for heavy-flavour decay muons at mid-rapidity in d-Au collisions at $\sqrt{s_{\rm{NN}}}=200$ GeV by the PHENIX experiment~\cite{Adare:2013lkk}, where the enhancement was not well described by models which include cold-nuclear matter effects. Model calculations at LHC energies are needed to interpret these measurements.  

\begin{figure}
\centering
\begin{subfigure}{0.45\textwidth}
\includegraphics[width=\textwidth]{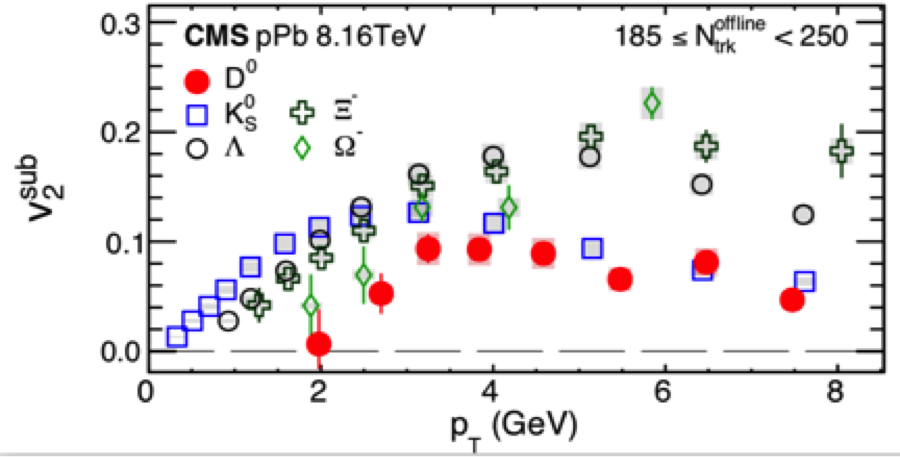}
\caption{}
\label{fig:cmspPbv2}
\end{subfigure}
\begin{subfigure}{0.45\textwidth}
\includegraphics[width=\textwidth]{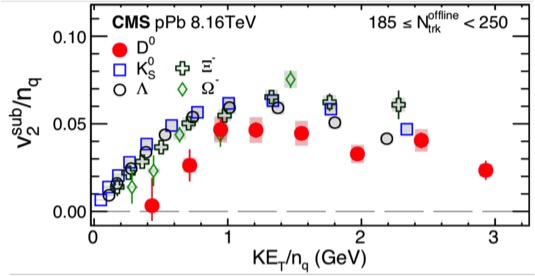}
\caption{}
\label{fig:cmspPbv2_KE}
\end{subfigure}
\caption{(a) $v_{2}$ vs. $p_{\rm{T}}$, (b) $v_2/n_q$ vs. $KE_{\rm{T}}/n_q$ for $\rm{D}^0$ mesons compared to different strange hadrons in high-multiplicity p-Pb collisions at  $\sqrt{s_{\rm{NN}}}=8.16$ TeV. }
\end{figure}

Two particle angular correlation measurements of light-flavour particles in p-A/d-A collisions have  shown long-range ridge structure and positive $v_{2}$ coefficients both at RHIC and LHC energies. To shed light on this unexpected observation, the $v_{2}$ of heavy-flavour particles was measured  by the PHENIX experiment at RHIC in high multiplicity d-Au collisions at $\sqrt{s_{\rm{NN}}}=200$ GeV and by ALICE~\cite{Acharya:2018dxy}, ATLAS and CMS~\cite{Sirunyan:2018toe} experiments at the LHC, in high-multiplicity p-Pb collisions at $\sqrt{s_{\rm{NN}}}= 5.02$ and $8.16$ TeV. In Figure~\ref{fig:cmspPbv2}, the $v_{2}$ of $\rm{D}^{0}$ mesons compared to that for various strange- hadrons in high-multiplicity p-Pb collisions is presented. A clear mass ordering is observed at low $p_{\rm{T}}$ in high-multiplicity p-Pb collisions, similar to that observed in A-A collisions~\cite{Sirunyan:2018toe}. To study the $v_{2}$ at the partonic level, the $v_{2}/n_{q}$ vs. $KE_{\rm{T}}/n_{q}$ (where $KE_{\rm{T}} = \sqrt{m^{2}+p_{\rm{T}}^2}-m$) was also obtained as shown in Figure~\ref{fig:cmspPbv2_KE}, where at low $KE_{\rm{T}}/n_{q}$, a universal trend for non-charm particles is observed while charm quarks have lower values unlike in Pb-Pb collisions~\cite{Sirunyan:2018toe}. This could indicate different initial- and final- state effects playing a role in small and large systems. These heavy-flavour measurements could  constraint the understanding of different possible underlying effects. 

\subsection*{Heavy-flavour hadronization}
\begin{figure}
	\centering
\begin{subfigure}{0.38\textwidth}
\includegraphics[width=\textwidth]{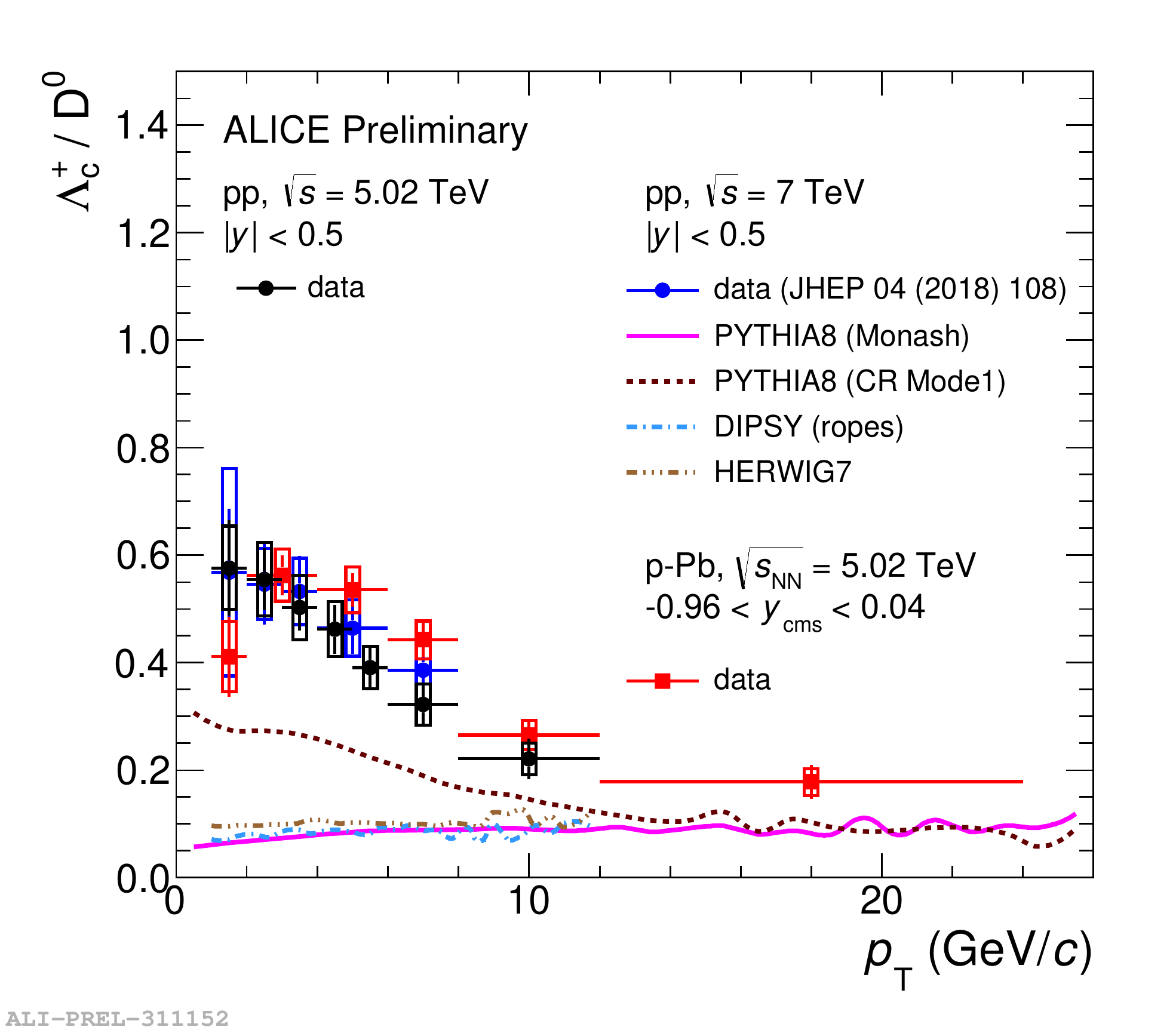}
\caption{}
\label{fig:Lambdac_pp_pPb}
\end{subfigure}
\begin{subfigure}{0.38\textwidth}
\includegraphics[width=\textwidth]{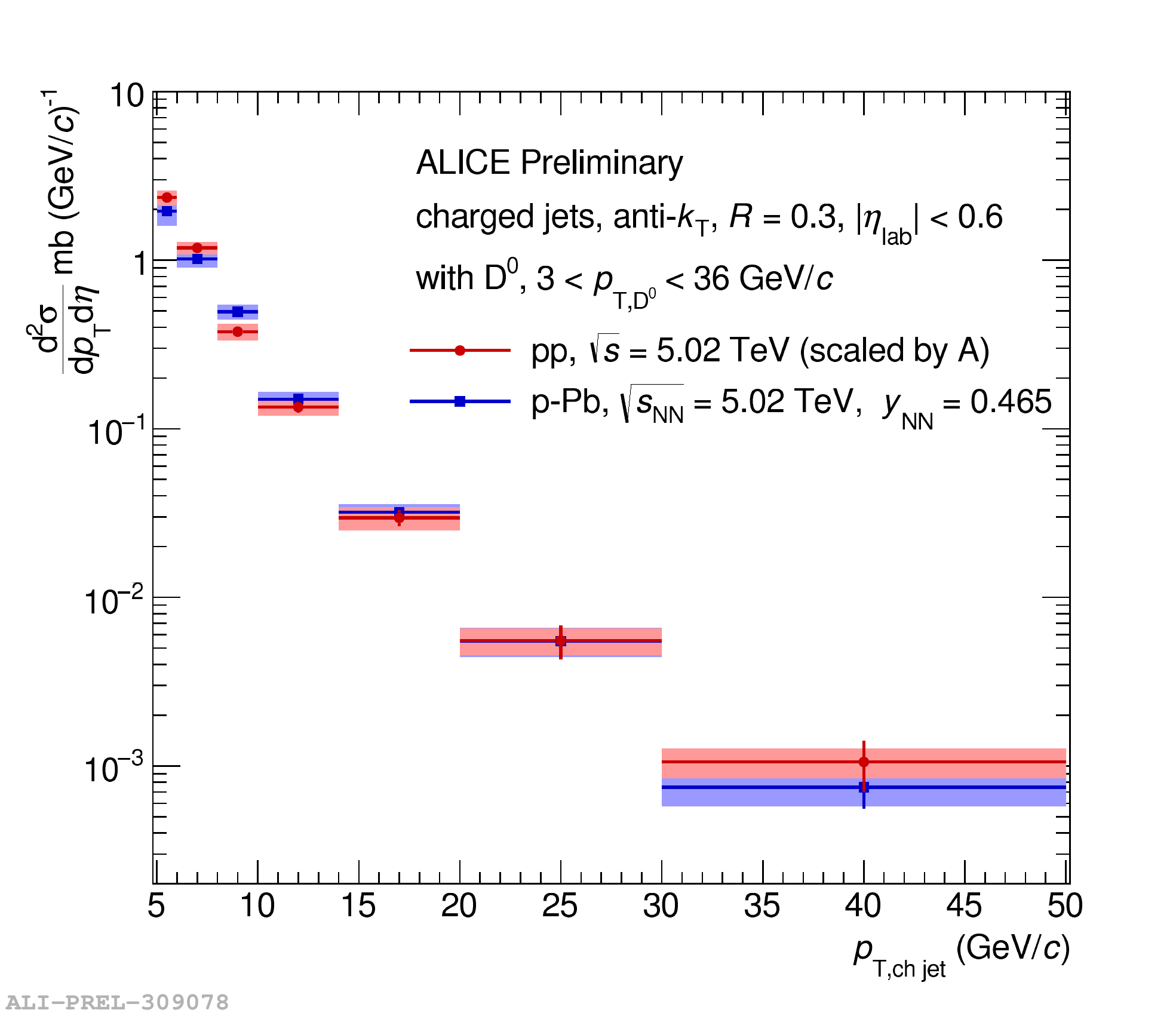}
\caption{}
\label{fig:DjetpPb}
\end{subfigure}
\caption{(a) $\Lambda_c^+/\rm{D}^0$ ratio as a function of $p_{\rm{T}}$ in pp collisions at $\sqrt{s}=5.02$ and 7 TeV and in p-Pb collisions at $\sqrt{s_{\rm{NN}}}=5.02$ TeV. (b) $\rm{D}^0$-jet cross-section with $3 < p_{\rm{T}}^{\rm{D}^{0}} < 36$ GeV/$c$ in pp and p-Pb collisions at $5.02$ TeV.}
\end{figure}

The study of the heavy-flavour hadronization process into baryons was performed by measuring the production of $\Lambda_c^+$ in pp collisions. The effects of cold nuclear matter on its production was studied with the measurement in p-Pb collisions. Figure~\ref{fig:Lambdac_pp_pPb} presents the $\Lambda_c^+/\rm{D}^{0}$ ratio measured in pp collisions at $\sqrt{s}=5.02$ and 7 TeV compared to the ratio in p-Pb collisions at  $\sqrt{s_{\rm{NN}}}=5.02$~TeV from ALICE. The $\Lambda_c^+/\rm{D}^{0}$ ratio versus transverse momentum measured in p-Pb collisions is consistent with that measured in pp collisions. The measurement was compared to models including PYTHIA8~\cite{Biro:1984cf}, DIPSY~\cite{Flensburg:2011kk} and HERWIG~\cite{Bahr:2008pv}, where all the models underestimate the data, while PYTHIA8 with enhanced color re-connection is closer to the measured values.

\subsection*{Heavy-flavour jet production}
Heavy-flavour tagged jet production in pp and p-Pb collisions was studied by measuring charged-jets containing heavy-flavour decay electrons or D mesons by ALICE. 
Figure~\ref{fig:DjetpPb} shows the $p_{\rm{T}}$-differential cross-section of jets containing $\rm{D}^0$ mesons with  $3 < p_{\rm{T}}^{\rm{D}^{0}} < 36$ GeV/$c$ in the transverse momentum range of $5 < p_{\rm{T}}^{\rm{ch,jet}} < 50$ GeV/$c$ in pp and p-Pb collisions. The cross-sections in pp collisions at $\sqrt{s}=5.02$ TeV and in p-Pb collisions at $\sqrt{s_{\rm{NN}}}=5.02$ TeV are consistent with each other and are in good agreement with NLO pQCD POWHEG+PYTHIA6 predictions. The $R_{\rm{pPb}}$ of jets with $\rm{D}^{0}$ meson in p-Pb collisions was shown to be consistent with unity. 

\subsection*{Beauty production}
\begin{figure}
	\centering
\begin{subfigure}{0.35\textwidth}
\includegraphics[width=\textwidth]{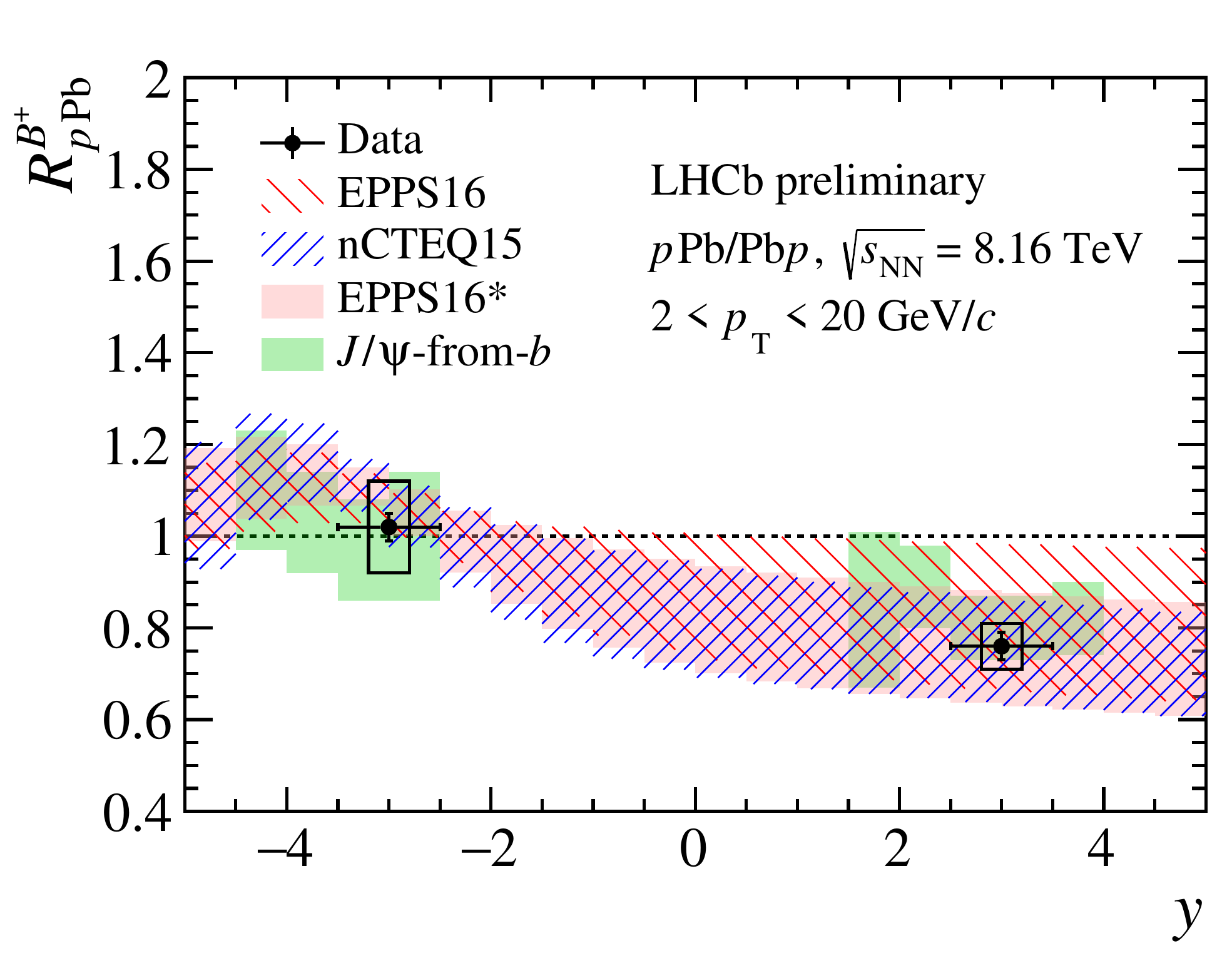}
\caption{}
\label{fig:BRpPb}
\end{subfigure}
\begin{subfigure}{0.35\textwidth}
\includegraphics[width=\textwidth]{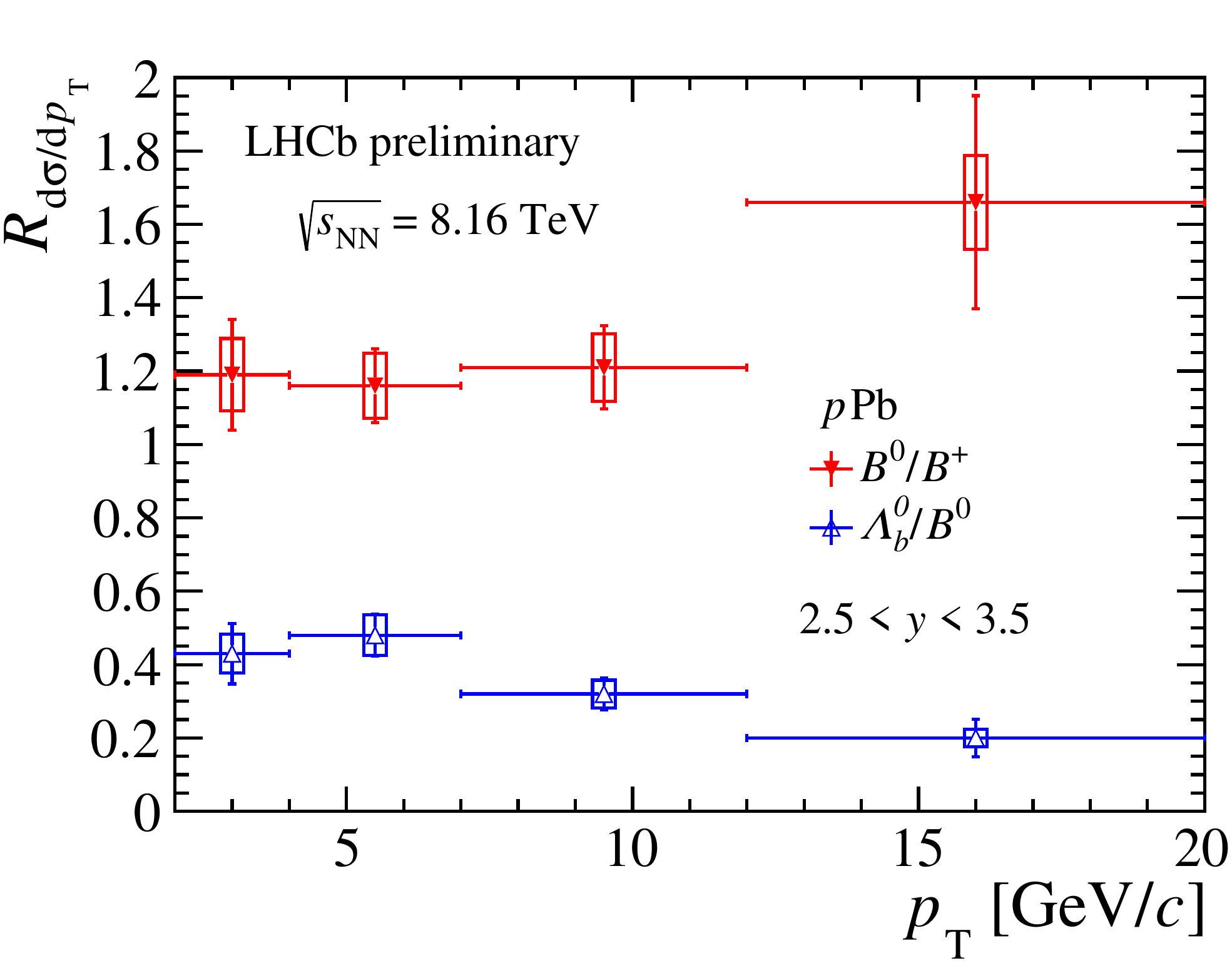}
\caption{}
\label{fig:Lambdab}
\end{subfigure}
\caption{(a) $R_{\rm{pPb}}$ of $\rm{B}^+$ mesons vs. rapidity for $2 < p_{\rm{T}} < 20$ GeV/$c$, (b) Ratio of $\rm{B}^{0}/\rm{B}^{+}$ and  $\Lambda^{0}_{b}/\rm{B}^{0}$ cross-sections vs. $p_{\rm{T}}$ for $2.5<y<3.5$ in p-Pb collisions at $\sqrt{s_{\rm{NN}}}=5.02$ TeV.}
\end{figure}

Measurements of beauty-quark production in p-Pb collisions was presented by LHCb using fully reconstructed b-hadrons, to study the modification of its production due to cold nuclear matter effects. The measurement was performed to $p_{\rm{T}}$ lower than the hadron mass, to help constrain the gluon wave function in the nucleus in the small $x$ region, where $x$ is the fraction of the nucleon momentum carried by the interacting gluon. The Figure~\ref{fig:BRpPb} presents the nuclear modification factor ($R_{\rm{pPb}}$) of $\rm{B}^{+}$ mesons  measured as a function of rapidity in p-Pb collisions for $2 < p_{\rm{T}} < 20$ GeV/$c$. A significant suppression in p-Pb collisions w.r.t to pp collisions is observed at forward rapidity while $R_{\rm{pPb}}$ is consistent with unity at backward rapidity. The study of b-quark fragmentation in p-Pb collisions was also performed by LHCb by measuring the ratio of the cross-section of $\rm{B}^{0}/\rm{B}^{+}$ and  $\Lambda^{0}_{b}/\rm{B}^{0}$. Shown in Figure~\ref{fig:Lambdab} is the ratio of the cross-section as a function of $p_{\rm{T}}$ for $2.5 < y<3.5$. The ratio of $\rm{B}^{0}/\rm{B}^{+}$ was shown to be independent in $p_{\rm{T}}$ and in rapidity
. The ratio of $\Lambda^{0}_{b}/\rm{B}^{0}$ exhibits a decreasing trend in $p_{\rm{T}}$ while showing no rapidity dependence~\cite{LHCb:2018jln}
. The ratio of the cross-sections of $\rm{B}^{0}/\rm{B}^{+}$ and $\Lambda^{0}_{b}/\rm{B}^{0}$ in p-Pb collisions are similar to that measured in pp collisions~\cite{Aaij:2013qqa,Aaij:2014jyk}.

\section{Heavy-flavour production in A-A collisions}
\subsection*{Beauty production}
\begin{figure}
	\centering
\begin{subfigure}{0.37\textwidth}
\includegraphics[width=\textwidth]{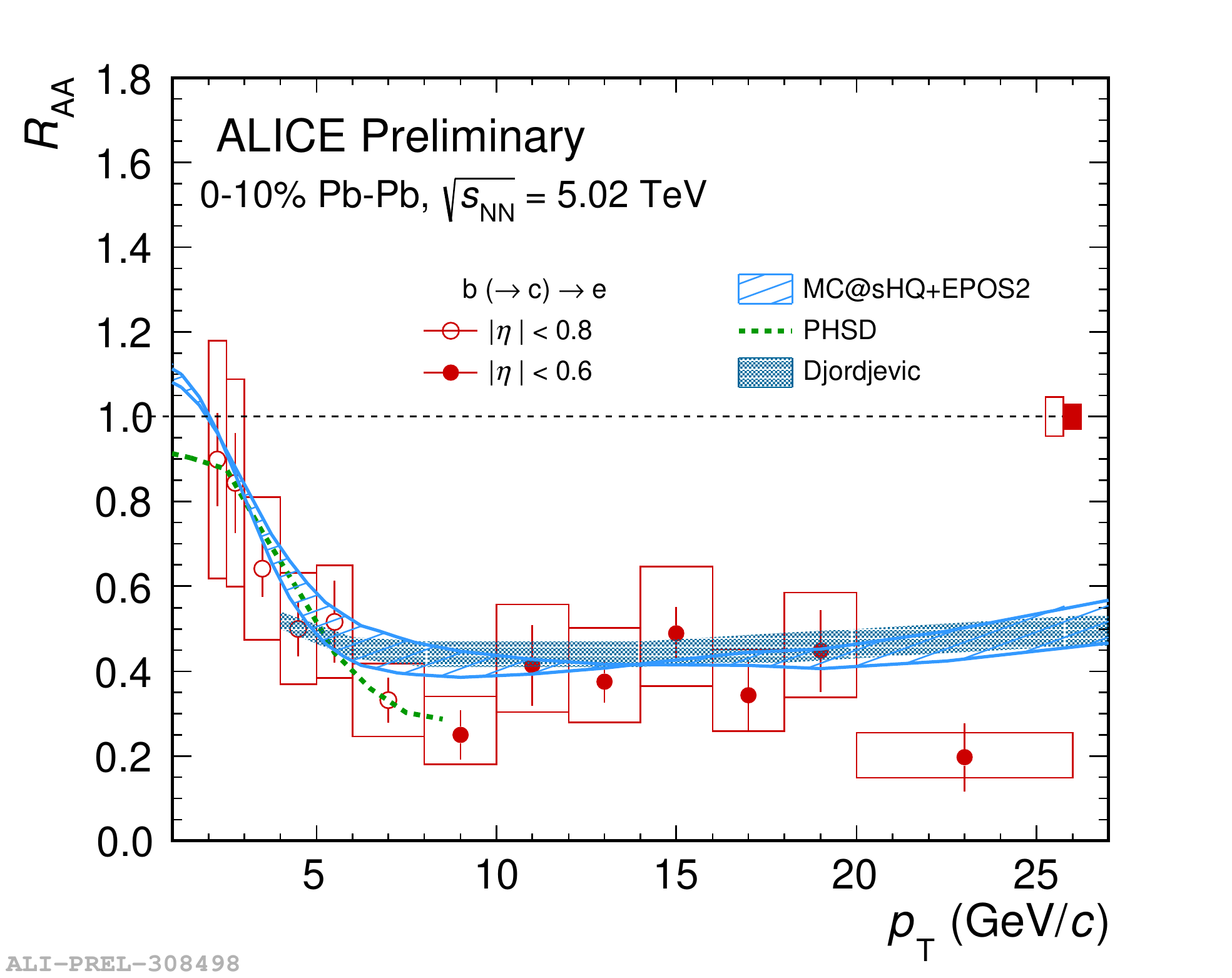}
\caption{}
\label{fig:bePb}
\end{subfigure}
\begin{subfigure}{0.33\textwidth}
\includegraphics[width=\textwidth]{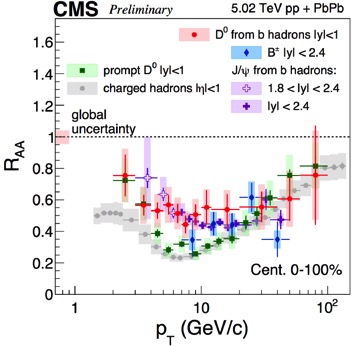}
\caption{}
\label{fig:Bcms}
\end{subfigure}
\caption{(a) $R_{\rm{AA}}$ of electrons from beauty-hadron decays in $0-10\%$ Pb-Pb collisions compared to model predictions. (b) $R_{\rm{AA}}$ of $\rm{B}^+$, non-prompt $\rm{D}^0$, non-prompt J/$\Psi$ compared to prompt $\rm{D}^0$ in $0-100\%$ Pb-Pb collisions. }
\end{figure}

Beauty quarks are 5 times heavier than charm quarks and around 5000 times heavier than up and down quarks making them ideal for studying the mass dependent energy loss of partons in the QGP. The nuclear modification factor of beauty hadrons was studied using B mesons and their decay particles at RHIC and at the LHC in Au-Au and Pb-Pb collisions, respectively. The $R_{\rm{AA}}$ of beauty-decay electrons measured by ALICE in $0-10\%$ Pb-Pb collisions at $\sqrt{s_{\rm{NN}}}=5.02$ TeV, as shown in Figure~\ref{fig:bePb}, is compared to model predictions from MC@sHQ+EPOS2~\cite{Nahrgang:2013xaa}, PHSD~\cite{Song:2015ykw} and Djordjevic~\cite{Djordjevic:2015hra},  which include both collisional and radiative energy loss for beauty quarks. The measurement is in good agreement with models in the full $p_{\rm{T}}$ range explored by the models. The $R_{\rm{AA}}$ of $\rm{B}^+$, non-prompt $\rm{D}^0$, non-prompt J/$\Psi$ was measured by CMS in miminum-bias Pb-Pb collisions at $\sqrt{s_{\rm{NN}}}=5.02$~TeV, and compared to the $R_{\rm{AA}}$ of prompt $\rm{D}^0$ mesons as shown in Figure~\ref{fig:Bcms}. The measurement shows a hint of a higher $R_{\rm{AA}}$ for B mesons than that of $\rm{D}^{0}$ mesons for $p_{\rm{T}}$ up to 10-15~GeV/$c$.  
 
\subsection*{Heavy-flavour hadronization} 
\begin{figure}
	\centering
\begin{subfigure}{0.33\textwidth}
\includegraphics[width=\textwidth]{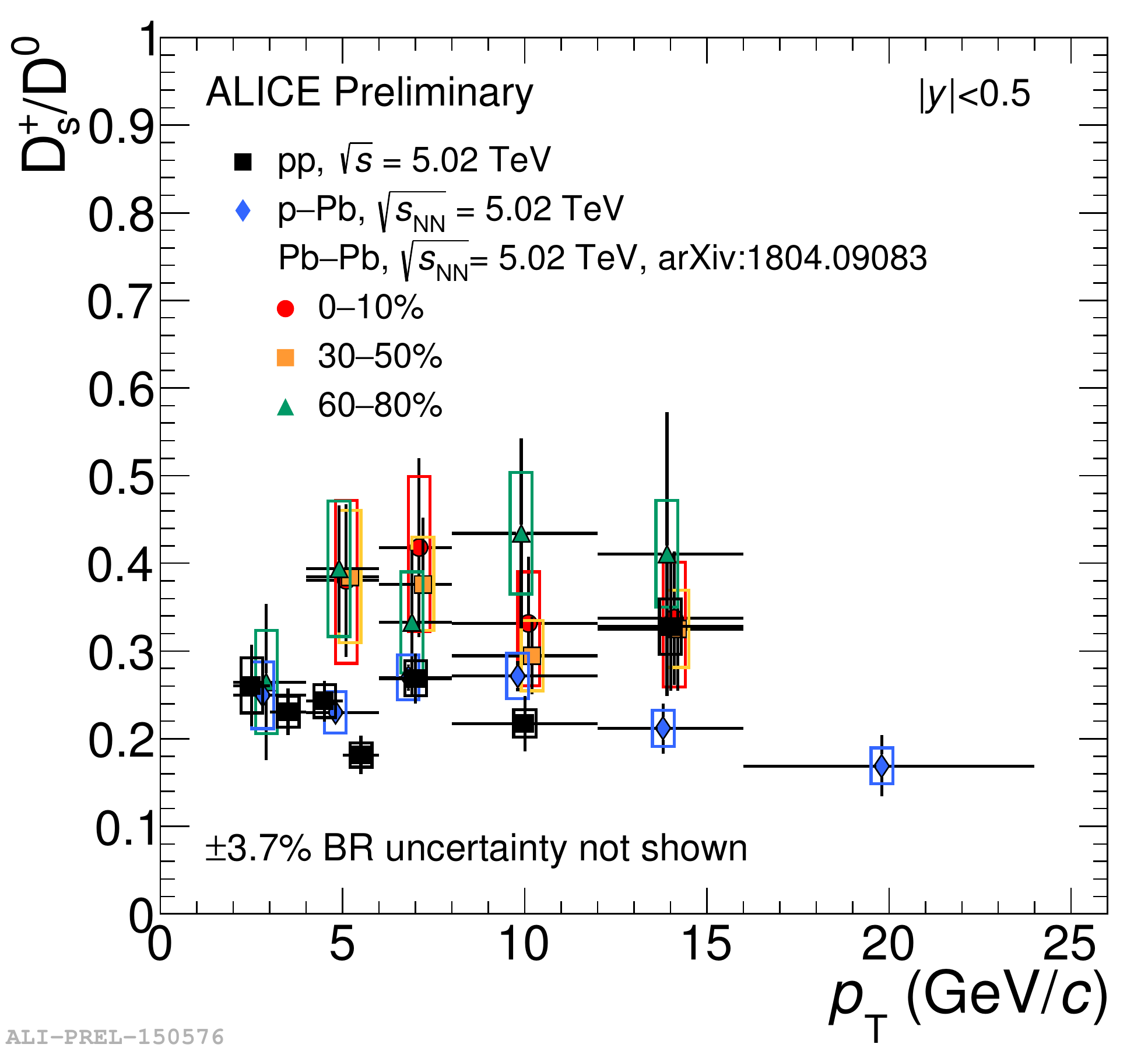}
\caption{}
\label{fig:DsPbPb}
\end{subfigure}
\begin{subfigure}{0.33\textwidth}
\includegraphics[width=\textwidth]{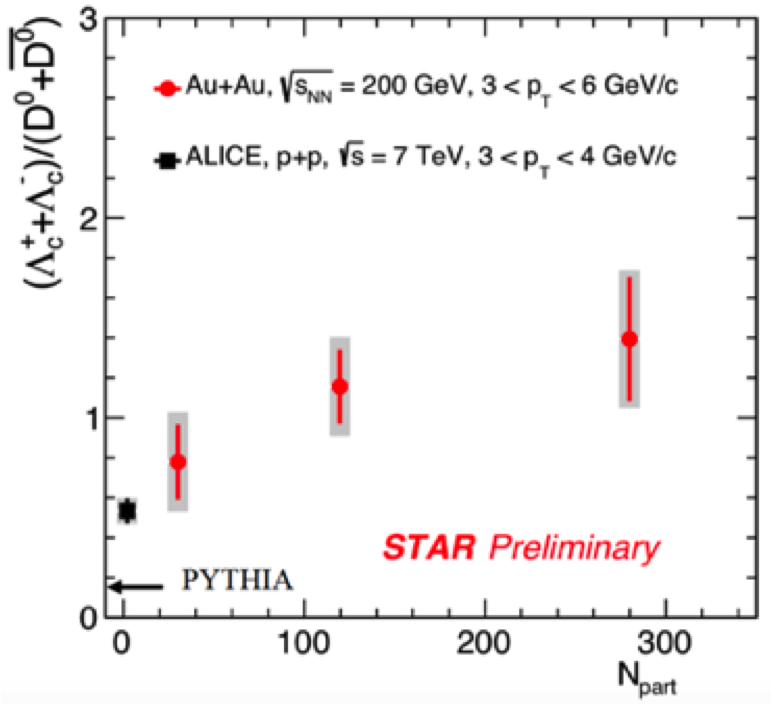}
\caption{}
\label{fig:LcAuAu}
\end{subfigure}
\caption{(a) $\rm{D}_s^+/\rm{D}^0$ ratio vs $p_{\rm{T}}$ in pp, p-Pb and in different centrality ranges in Pb-Pb collisions at 5.02 TeV. (b) $\Lambda_c/\rm{D}^0$ ratio vs number of participant nucleons in Au-Au collisions compared to the ratio in pp collisions, where the centrality of Au-Au collisions increases from left to right.}
\end{figure}

The hadronization mechanisms of heavy quarks was studied by measuring the production of $\rm{D}_s^+$ mesons by ALICE and STAR experiments. Models predict that low-momentum heavy quarks could hadronize not only via fragmentation in the vacuum, but also via  recombination or coalescence with other quarks in the medium. Due to the large abundance of strange quarks in A-A collisions relative to pp collisions, an increased production of $\rm{D}_s^+$ mesons relative to non-strange D~mesons is expected. The ratio of $\rm{D}_s^+$ to $\rm{D}^0$ mesons as a function of $p_{\rm{T}}$ in pp, p-Pb and in different centrality ranges in Pb-Pb collisions as measured by ALICE is shown in Figure~\ref{fig:DsPbPb}. 
The $\rm{D}_s^+/\rm{D}^0$ ratio indicates a larger value in Pb-Pb collisions than in pp collisions in all centrality classes, hinting at an enhancement of $\rm{D}_s^+$ meson production due to coalescence in Pb-Pb collisions. 

To further investigate the hadronization mechanisms,  heavy-flavour baryon production was studied, as the models that include coalescence predict an enhanced baryon-to-meson ratio at low and intermediate $p_{\rm{T}}$ in comparison to that expected in pp collisions. The STAR and ALICE collaboration measured the $\Lambda_c/\rm{D}^0$ ratios in Au-Au and Pb-Pb collisions, respectively. Figure~\ref{fig:LcAuAu} presents the $\Lambda_c/\rm{D}^0$ ratio measured in Au-Au collisions at $\sqrt{s_{\rm{NN}}}= 200$ GeV as a function of the number of participant nucleons, compared to the measured value in pp collisions at $\sqrt{s}=7$ TeV. The $\Lambda_c/\rm{D}^0$ ratio increases with centrality and its value in peripheral Au-Au collisions is close to that in pp collisions. These measurements indicate a significant enhancement of baryon over meson yield, which is closer to the predictions from models that include hadronization via coalescence. 

\subsection*{Heavy-flavour correlations and jets}
\begin{figure}
	\centering
\begin{subfigure}{0.32\textwidth}
\includegraphics[width=\textwidth]{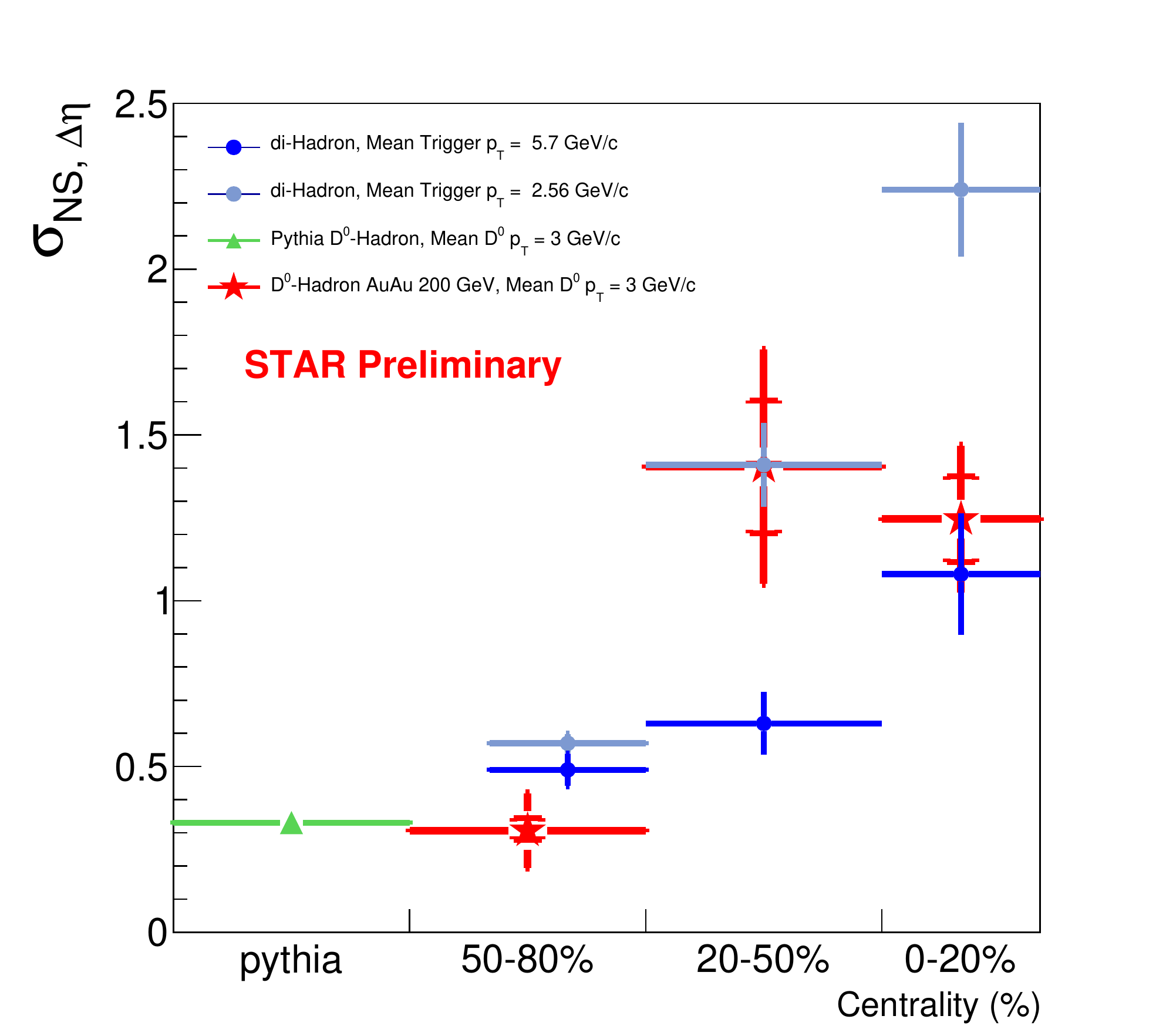}
\caption{}
\label{fig:widtheta}
\end{subfigure}
\begin{subfigure}{0.28\textwidth}
\includegraphics[width=\textwidth]{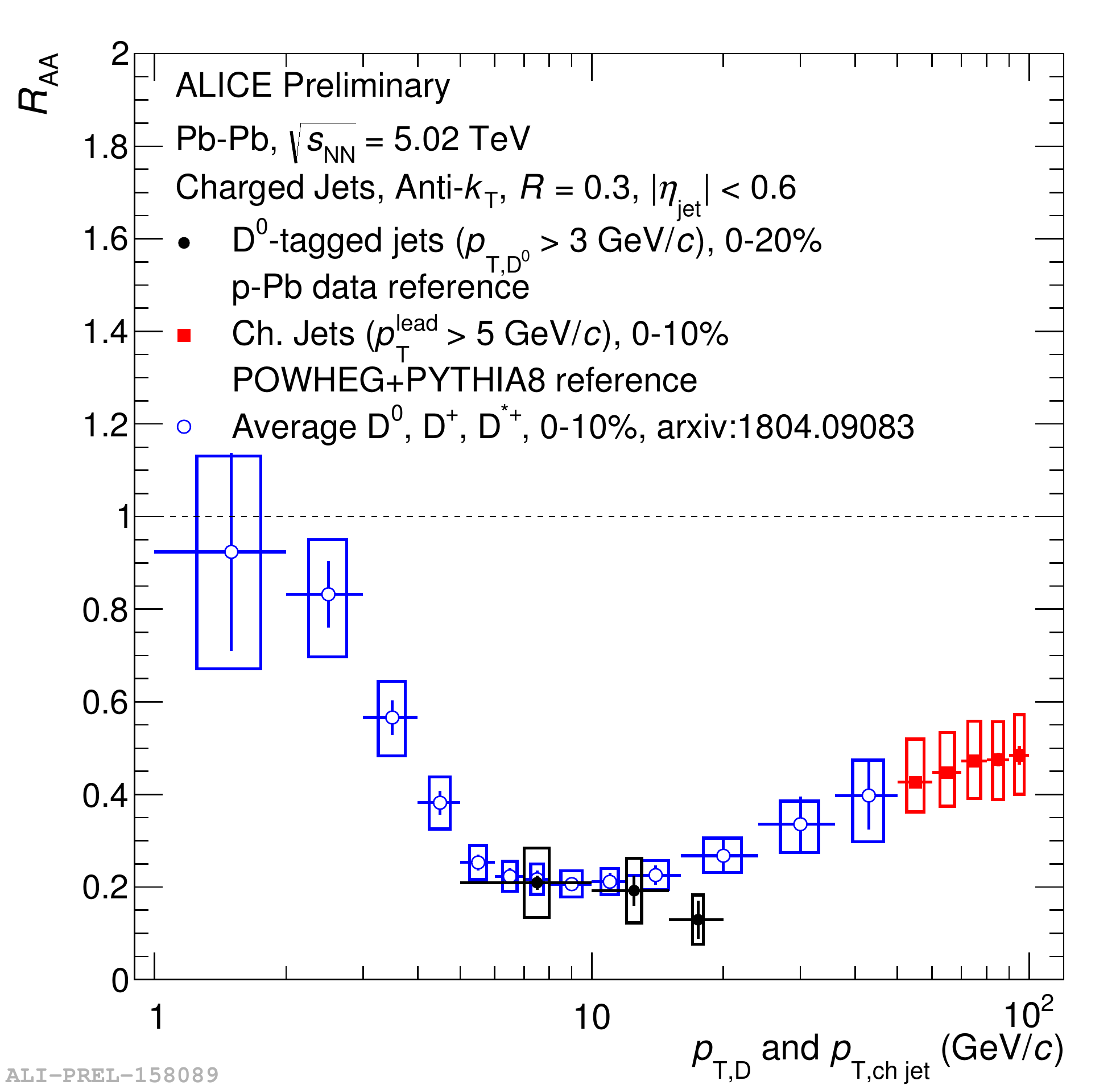}
\caption{}
\label{fig:D0jetRaa}
\end{subfigure}
\begin{subfigure}{0.28\textwidth}
\includegraphics[width=\textwidth]{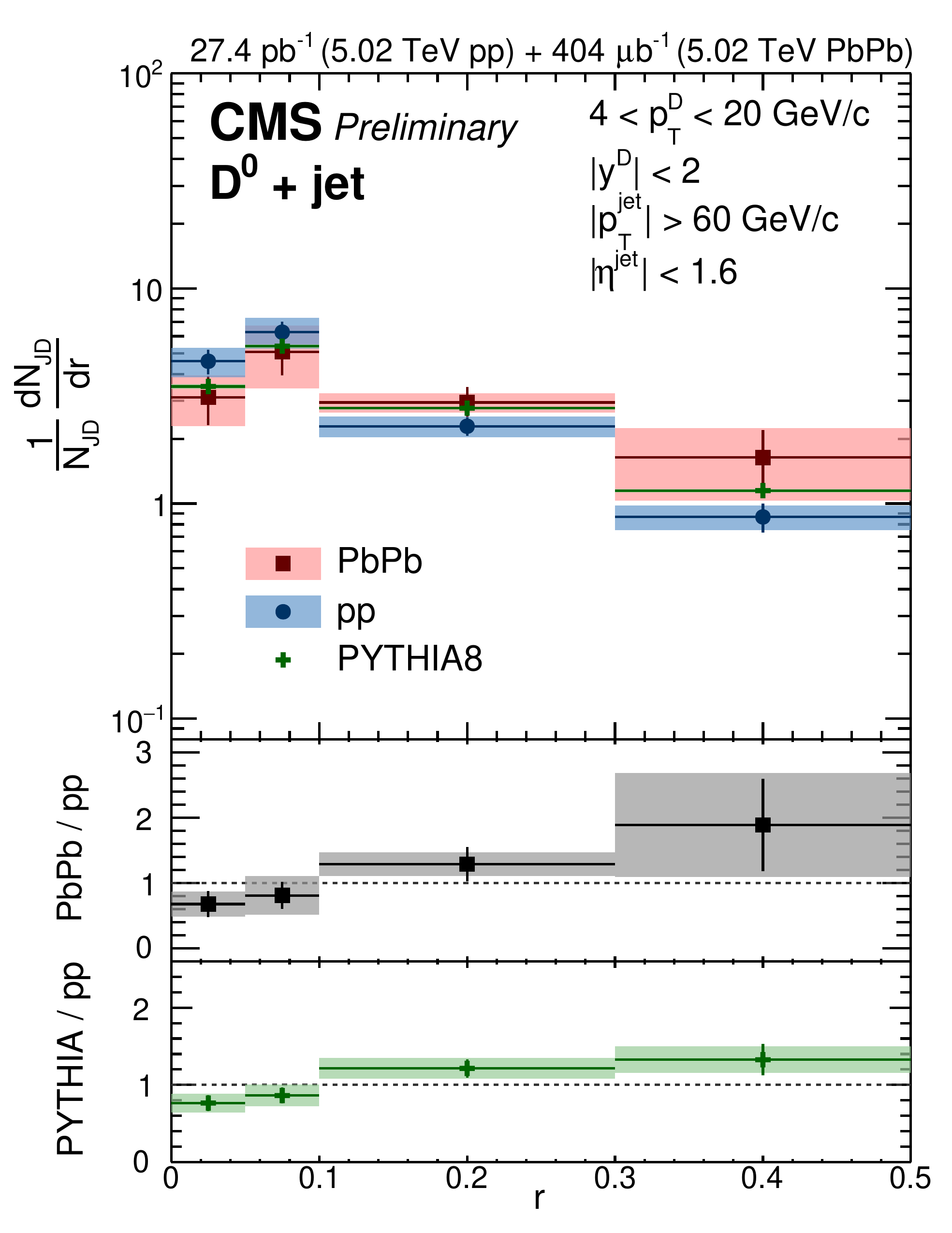}
\caption{}
\label{fig:D0radCMS}
\end{subfigure}
\caption{(a) Near-side width of angular correlations of $\rm{D}^0$ and charged particles along $\Delta\eta$ in different centrality ranges of Au-Au collisions compared to PYTHIA simulations. (b) $R_{\rm{AA}}$ of $\rm{D}^0$-tagged jets compared to $\rm{D}^0$ mesons and charged jets in Pb-Pb collisions. (c) Radial distribution of $\rm{D}^0$ mesons in jets in Pb-Pb collisions compared to pp collisions.}
\end{figure}

The heavy-flavour jet structure was studied by the STAR collaboration by measuring two particle $\Delta\varphi,\Delta\eta$ angular correlations of $\rm{D}^0$ mesons and charged particles in Au-Au collisions at $\sqrt{s_{\rm{NN}}}~=~200$~GeV in different centrality ranges. The 2-D correlation structure for $2<~p_{\rm{T}}^{\rm{D}^0}<10$~GeV/$c$ and $p_{\rm{T}}^{\rm{ch}} > 0.15$~GeV/$c$ was fit with near- and away-side Gaussian, a constant and $v_2$ terms to extract the near- and away-side jet properties. The near-side width along $\Delta\varphi$ and $\Delta\eta$ was extracted and compared to that obtained from light-flavour particle correlations in the same centrality bins and to a PYTHIA prediction. The width of the near-side correlation structure along $\Delta\eta$ is shown in Figure~\ref{fig:widtheta}. The measurement indicates an increase in the near-side width from peripheral to central Au-Au collisions, while the width in peripheral Au-Au collisions is consistent with that obtained from PYTHIA. The broadening of the correlation width versus centrality along $\Delta\eta$ was  observed to be larger than the broadening along $\Delta\varphi$. 

The modification of fragmentation of heavy-flavour jets was studied by ALICE with jets containing ${\rm{D}^0}$ mesons in Pb-Pb collisions at $\sqrt{s_{\rm{NN}}} = 5.02$~TeV for  $5< p_{\rm{T}}^{\rm{Jet}}< 20$~GeV/$c$ with $p_{\rm{T}}^{\rm{D^0}} > 3$~GeV/$c$~\cite{CMS:2018ovh}. Figure~\ref{fig:D0jetRaa} shows the $R_{\rm{AA}}$ of ${\rm{D}^0}$-tagged jets, where a strong suppression similar to that of prompt D mesons is observed. The structure of heavy-flavour jets and their modification in heavy-ion collisions was studied by CMS by comparing the measurement of the radial distribution of $\rm{D^0}$ mesons in jets in Pb-Pb collisions to that in pp collisions. Figure~\ref{fig:D0radCMS} shows the $\rm{D^0}$ radial distribution ($r$) within a jet in pp and Pb-Pb collisions for $4 < p_{\rm{T}}^{\rm{D^0}} < 20$ GeV/$c$; the ratio of the spectra in Pb-Pb to pp collisions is also presented in the lower panel of the same Figure. The ratio increases with $r$, indicating that $\rm{D^0}$ mesons at low $p_{\rm{T}}$ are farther away from the jet-axis in Pb-Pb collisions compared to pp collisions.

\subsection*{Directed flow}

\begin{figure}
	\centering
\begin{subfigure}{0.5\textwidth}
\includegraphics[width=\textwidth]{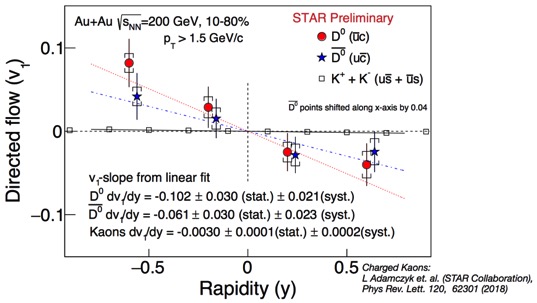}
\caption{}
\label{fig:v1star}
\end{subfigure}
\begin{subfigure}{0.35\textwidth}
\includegraphics[width=\textwidth]{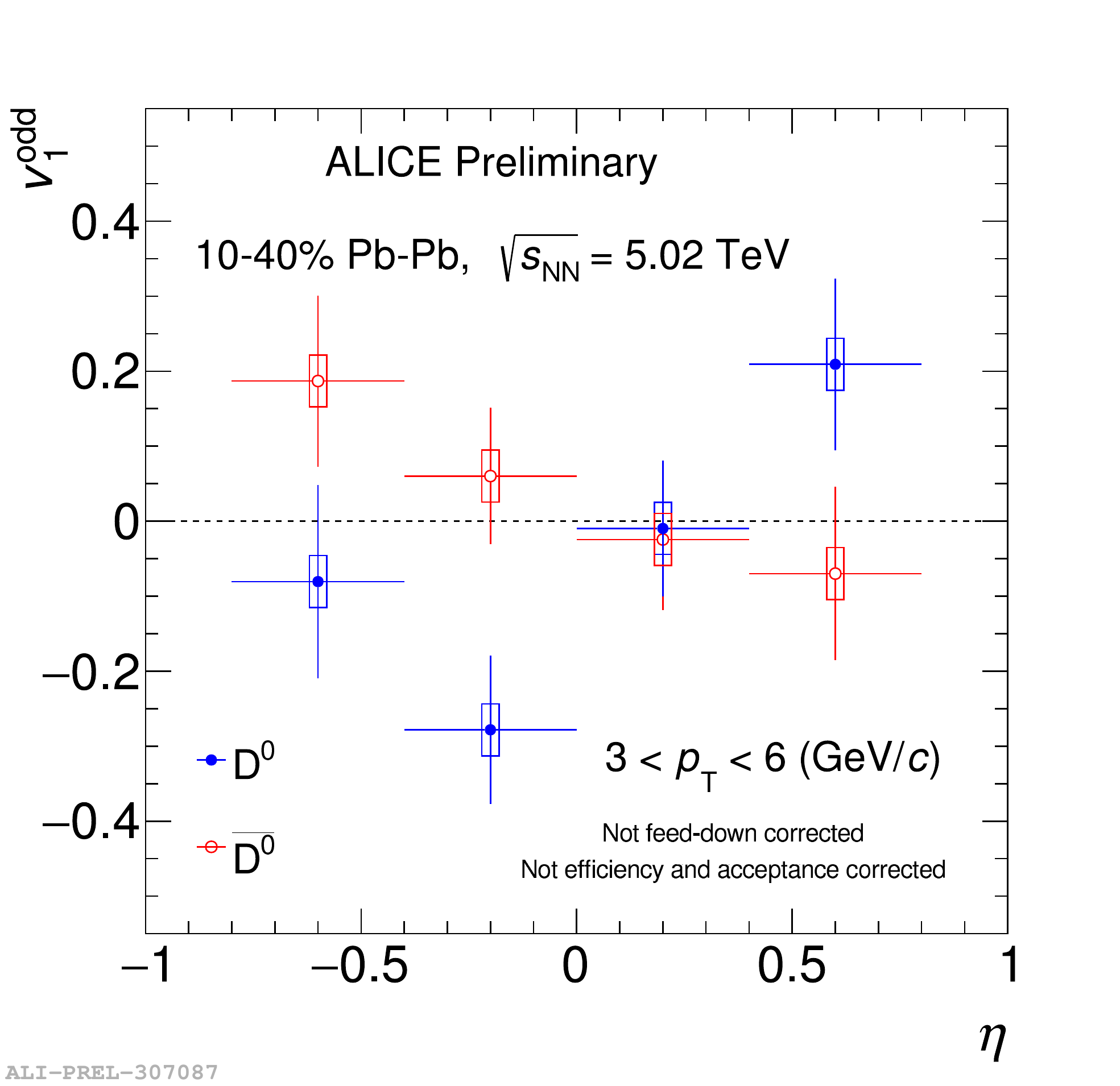}
\caption{}
\label{fig:v1alice}
\end{subfigure}
\caption{(a) Directed flow of $\rm{D^0}$ mesons vs. rapidity in $10-80\%$ central Au-Au collisions at $\sqrt{s_{\rm{NN}}} = 200$ GeV, (b)$~10-40\%$ central Pb-Pb collisions at $\sqrt{s_{\rm{NN}}} = 5.02$ TeV. }
\end{figure}

Charge-dependent directed flow ($v_1$) of heavy quarks is a good probe to study the dynamical effects of the electro-magnetic field produced by the charged-spectator nucleons in non-central A-A collisions. STAR and ALICE experiments observed the directed flow of $\rm{D^0}$ mesons as a function of rapidity in Au-Au at $\sqrt{s_{\rm{NN}}} = 200$ GeV and in Pb-Pb collisions at $\sqrt{s_{\rm{NN}}} = 5.02$ TeV, respectively. The rapidity-odd components of $v_1$ for $\rm{D^0}$ and $\overline{\rm{D}^0}$ were measured to have non-zero values in Au-Au collisions, as shown in Figure~\ref{fig:v1star}, with the slopes for $\rm{D^0}$ mesons greater than for the non-charm mesons (K). At the LHC the rapidity-odd component of $v_1$, measured in Pb-Pb collisions, shows an opposite trend versus rapidity for  $\rm{D^0}$ and $\overline{\rm{D}^0}$ but with large uncertainties, as shown in Figure~\ref{fig:v1alice}. Model predictions are required to gain insight into these measurements. The precision of the data is also expected to be improved with the new Pb-Pb data collected at the LHC in 2018. 

\section{Summary}
In this paper I have summarized some of the new and exciting measurements of open heavy-flavour particles that were presented by experiments at RHIC and at the LHC at the Hard Probes 2018 conference. For p-A collisions, while many measurements can be described by models that include cold-nuclear matter effects, a better understanding of  initial- and final-state effects is required to describe the data measured in high-multiplicity events. In A-A collisions, with high-precision D meson measurements available, the community is extending the study of transport properties of the QGP by using beauty quarks as probes. Hadronization mechanisms are being extensively studied using heavy-flavour baryons and strange heavy-flavour mesons. New measurements of heavy-flavour correlations and jets were performed to study the medium modification of jet-fragmentation and structure. With planned new experiments in the coming years, including sPHENIX at RHIC and upgrades of the current experiments in ALICE and LHCb at the LHC, we can look forward to having more extensive and precise heavy-flavour measurements in the near future.

\end{document}